\documentclass[pra,showpacs,twocolumn]{revtex4}
\usepackage{bm,graphicx,amsmath,amssymb}
\usepackage{epsf,graphics,graphicx}
\usepackage{float}
\usepackage{subfigure}
\usepackage{color}
\usepackage[normalem]{ulem}
\usepackage[caption=false]{subfig}

\newcommand{\ket}[1]{|#1\rangle}

\begin{document}
\title{Entanglement duality in spin-spin interactions}
\author{Vahid Azimi Mousolou\footnote{Electronic address: v.azimi@sci.ui.ac.ir}}
\affiliation{Department of Applied Mathematics and Computer Science, 
Faculty of Mathematics and Statistics, 
University of Isfahan, Isfahan 81746-73441, Iran}
\affiliation{Department of Physics and Astronomy, Uppsala University, Box 516, 
SE-751 20 Uppsala, Sweden}

\author{Anders Bergman}
\affiliation{Department of Physics and Astronomy, Uppsala University, Box 516, 
SE-751 20 Uppsala, Sweden}

\author{Anna Delin}
\affiliation{Department of Physics and Astronomy, Uppsala University, Box 516, 
SE-751 20 Uppsala, Sweden}
\affiliation{Department of Applied Physics, School of Engineering Sciences, 
KTH Royal Institute of Technology, AlbaNova University Center, SE-10691 Stockholm, 
Sweden}
\affiliation{Swedish e-Science Research Center (SeRC), KTH Royal Institute of Technology, 
SE-10044 Stockholm, Sweden}

\author{Olle Eriksson}
\affiliation{Department of Physics and Astronomy, Uppsala University, Box 516, 
SE-751 20 Uppsala, Sweden}
\affiliation{School of Science and Technology, \"Orebro University, SE-701 82, 
\"Orebro, Sweden}

\author{Manuel Pereiro}
\affiliation{Department of Physics and Astronomy, Uppsala University, Box 516, 
SE-751 20 Uppsala, Sweden}

\author{Danny Thonig}
\affiliation{School of Science and Technology, \"Orebro University, SE-701 82, 
\"Orebro, Sweden}

\author{Erik Sj\"oqvist\footnote{Electronic address: 
erik.sjoqvist@physics.uu.se}}
\affiliation{Department of Physics and Astronomy, Uppsala University, 
Box 516, SE-751 20 Uppsala, Sweden}
\date{\today}
\begin{abstract}
We examine entanglement of thermal states for spin-$\frac{1}{2}$ dimers in external magnetic 
fields. Entanglement transition in the temperature-magnetic field plane demonstrates a duality 
in spin-spin interactions. This identifies dual categories of symmetric and anti-symmetric 
dimers with each category classified into toric entanglement classes. The entanglement 
transition line is preserved from each toric entanglement class to its dual toric class. The 
toric classification is an indication of topological nature of the entanglement.  
\end{abstract}
\pacs{}
\maketitle
\section{Introduction}

The classification concept has been incorporated into different fields of studies ranging from 
biological systems to abstract topology, in order to categorize relevant objects based on 
shared characteristics. Scientific classification schemes have not only led to new discoveries 
of materials and resources, such as in the topological classification of matter 
\cite{chiu16} and of entangled quantum states \cite{johansson14},  but have also significantly 
helped to find the most efficient and robust 
approaches to technological advances.   

One of the fundamental characteristics of quantum mechanics is the quantum correlation 
or the entanglement, which lies at the heart of the difference between the classical and 
quantum world \cite{einstein1935, schrodinger1935, bell1964}. It is widely believed to be much 
stronger than classical correlations and over the years has become a critically important 
resource for many applications in quantum technology including quantum computing, 
quantum cryptography, quantum communication, and hyper-sensitive measurements 
\cite{nielsen2010}. As a primary attribute of quantum mechanics, quantum entanglement 
is also much more involved with the foundations, predictions and interpretations of quantum 
phenomena. It is strongly linked to the concepts of quantum phase transition and quantum 
geometric phases \cite{osterloh2002, osborne2002, wei2005, orus2008, son2011, azimi-mousolou2013}. Nonetheless, the mysterious characteristics of the quantum entanglement 
still needs to be explored in-depth.

Here we use entanglement in terms of concurrence \cite{hill1997,wootters98} from a 
new perspective, i.e. classification, in order to analyze spin-spin interactions in dimers and classify 
them into entanglement classes. For this, we focus on the entanglement transition line for 
thermal state of the general traceless spin-pair model in the external parameter space 
specified by temperature and applied magnetic field. With this we categorize dimers into 
dual categories of symmetric and anti-symmetric dimers. Each category is classified into 
toric entanglement classes, where each class together with its dual are distinguished by 
the same entanglement transition line in temperature-magnetic field plane. As a nontrivial 
example, we introduce dual symmetric and anti-symmetric Heisenberg spin-pair interactions 
and specify their toric entanglement classes. We note that in Ref. \cite{arnesen01}, the 
entanglement between two spins in a one dimensional Heisenberg chain has also been 
studied as a function of temperature and external magnetic field, but not from the classification 
approach, which is the main concern here. 

\section{General Model Hamiltonian}
We start with the general traceless spin-pair model described by the Hamiltonian ($\hbar = 1$ 
throughout the paper)
\begin{eqnarray}
H(\omega,\mathbb{J}) = \frac{\omega_+}{2} \sigma_z^{(1)} + \frac{\omega_-}{2} \sigma_z^{(2)} + 
\boldsymbol{\sigma}^{(1)} \cdot \mathbb{J} \boldsymbol{\sigma}^{(2)} , 
\label{eq:Hamiltonian}
\end{eqnarray}
with the real-valued matrix 
\begin{eqnarray}
\mathbb{J} = \left( \begin{array}{ccc} J_{xx} & J_{xy} & 0 \\ 
J_{yx} & J_{yy} & 0 \\  
0 & 0 & J_{zz} \end{array} \right),
\end{eqnarray}
where
\begin{eqnarray}
J_{xx} & = & \frac{1}{2} (J+r) , \ J_{yy} = \frac{1}{2} (J-r) , 
\nonumber \\ 
J_{xy} & = & \frac{1}{2} \left( K - D \right), \ J_{yx} = \frac{1}{2} \left( K + D \right) , 
\nonumber \\ 
\omega_{\pm} & = & \omega \pm \Delta.
\end{eqnarray}
This model accounts for a wide range of important spin-spin interaction systems, including 
Heisenberg ($J=J_{zz}$), Dzyaloshinskii–Moriya (D), and symmetric exchange ($K$), as 
well as XY anisotropy ($r$). It further allows for the spins to interact differently with 
the external magetic field ($\Delta$), due to field inhomogenties or differences in magnetic 
moments.

The two-qubit Hamiltonian is chosen so as to ensures that the thermal equilibrium state at 
temperature $T$ is of a $X$-type:
\begin{eqnarray}
\varrho_T =  \frac{1}{\mathcal{Z}} e^{-H/T} = \left( \begin{array}{cccc} 
\rho_{11} & 0 & 0 & \rho_{14} \\
0 & \rho_{22} & \rho_{23} & 0 \\
0 & \rho_{32} & \rho_{33} & 0 \\
\rho_{41} & 0 & 0 & \rho_{44} \end{array} \right).
\end{eqnarray}
in the ordered product qubit-qubit basis $\{ \ket{00}, \ket{01}, \ket{10}, \ket{11} \}$ and 
we put Boltzmann's constant $k_B=1$ from now. Explicitly, we have 
\begin{eqnarray}
\rho_{11}  & = &\frac{1}{\mathcal{Z}}e^{-\frac{J_{zz}}{T}}[\cosh\frac{\epsilon_1}{T} - 
\sinh\frac{\epsilon_1}{T}\cos\vartheta], 
\nonumber \\ 
\rho_{44}  & = &\frac{1}{\mathcal{Z}}e^{-\frac{J_{zz}}{T}}[\cosh\frac{\epsilon_1}{T} + 
\sinh\frac{\epsilon_1}{T}\cos\vartheta], 
\nonumber \\ 
\rho_{14} & = &\rho_{41}^{*} =-\frac{1}{\mathcal{Z}}e^{-\frac{J_{zz}}{T}} 
e^{-i\varphi}\sinh\frac{\epsilon_1}{T}\sin\vartheta,
\nonumber \\ 
\rho_{22}  & = &\frac{1}{\mathcal{Z}}e^{\frac{J_{zz}}{T}}[\cosh\frac{\epsilon_2}{T} - 
\sinh\frac{\epsilon_2}{T}\cos\theta], 
\nonumber \\ 
\rho_{33}  & = &\frac{1}{\mathcal{Z}}e^{\frac{J_{zz}}{T}}[\cosh\frac{\epsilon_2}{T} + 
\sinh\frac{\epsilon_2}{T}\cos\theta], 
\nonumber \\ 
\rho_{23} & = &\rho_{32}^{*} =-\frac{1}{\mathcal{Z}}e^{\frac{J_{zz}}{T}} 
e^{-i\phi}\sinh\frac{\epsilon_2}{T}\sin\theta,
\end{eqnarray}
where 
\begin{eqnarray}
\tan \vartheta & = & \frac{\sqrt{r^2 + K^2}}{\omega} ,\ \ \ \ \ \ \ \  
\tan \varphi = \frac{K}{r} , \\ 
\tan \theta & = & \frac{\sqrt{J^2 + D^2}}{\Delta}, \ \ \ \ \ \ \ \  
\tan \phi = \frac{D}{J} ,
\nonumber \\ 
\epsilon_1& = &\sqrt{\omega^2 + r^2 + K^2},\ \ \ \ \ \ \epsilon_2 = 
\sqrt{\Delta^2 + J^2+D^2},\ \ 
\nonumber
\end{eqnarray}
and $\mathcal{Z}=2\left(e^{-\frac{J_{zz}}{T}}\cosh\frac{\epsilon_1}{T} + 
e^{\frac{J_{zz}}{T}}\cosh\frac{\epsilon_2}{T}\right)$ the partition function.

The above $X$-state form is suitable in our analysis form two points of views: (i) as pointed 
out before, it accounts for thermal states of several important spin-spin interaction models, 
and (ii) the entanglement measure concurrence $C(\varrho_T)$ \cite{hill1997,wootters98} 
can be calculated analytically. Indeed, one finds \cite{mazzola10}
\begin{eqnarray}
C(\varrho_T) = 2\max \big\{C_1, C_2 ,0 \big\},
\label{eq:concurrence_general}
\end{eqnarray}
where
\begin{eqnarray}
C_1&=& |\rho_{14}| - \sqrt{\rho_{22} \rho_{33}} , \nonumber \\ 
C_2&=&|\rho_{23}| - \sqrt{\rho_{11} \rho_{44}}.
\label{eq:concurrence_elements}
\end{eqnarray}

We pursue by exploring the entanglement transition, for which one may solve 
\begin{eqnarray}
\max \big\{C_1, C_2  \big\}=0
\label{eq:concurrence_critical}
\end{eqnarray}
to extract the critical line in the temperature-magnetic field parameter space.
Eq.~\eqref{eq:concurrence_critical} gives rise to the following duality 
\begin{eqnarray}
(\text{I})\ \ \ \ \ C_2\le C_1&=&0\nonumber\\
(\text{II})\ \ \ \ \ C_1\le C_2&=&0,
\end{eqnarray}
which becomes 
\begin{eqnarray}
(\text{I})\ \ \ \ \ e^{-\frac{2J_{zz}}{T}}p^2-e^{\frac{2J_{zz}}{T}}q^2&=&e^{\frac{2J_{zz}}{T}}\nonumber\\
(\text{II})\ \ \ \ \ e^{\frac{2J_{zz}}{T}}q^2-e^{-\frac{2J_{zz}}{T}}p^2&=&e^{-\frac{2J_{zz}}{T}},
\label{EDE}
\end{eqnarray}
for our model system with $p=f(\epsilon_1,\nu)$ and $q=f(\epsilon_2,\theta)$ given by the 
function $f(x,y)=\sinh\frac{x}{T}\sin y$. 

This duality equation allows us to categorize the spin-spin interactions in dimers into a 
dual categories of interactions. Here we focus on symmetric and antisymmetric dimers 
in the sense that the two spins of same size forming the dimer possess parallel or antiparallel 
orientations of the respective magnetic moments. For each case the corresponding Hamiltonian 
would have the following form 
\begin{itemize}
\item {\it Symmetric}: Here with parallel magnetic moments we have $\omega_+=\omega_-=B$ 
and therefore  
\begin{eqnarray}
H_{\text{s}} & = & \frac{B}{2} [\sigma_z^{(1)} + \sigma_z^{(2)}] + 
J_{zz}\sigma^{(1)}_{z}\sigma^{(2)}_{z} 
\nonumber\\
 & & + \frac{J}{2}[\sigma^{(1)}_{x}\sigma^{(2)}_{x} + \sigma^{(1)}_{y}\sigma^{(2)}_{y}] - 
 \frac{D}{2}[\sigma^{(1)}_{x}\sigma^{(2)}_{y}-\sigma^{(1)}_{y}\sigma^{(2)}_{x}] 
\nonumber\\
 & & + \frac{r}{2}[\sigma^{(1)}_{x}\sigma^{(2)}_{x}-\sigma^{(1)}_{y}\sigma^{(2)}_{y}]+\frac{K}{2}[\sigma^{(1)}_{x}\sigma^{(2)}_{y}+\sigma^{(1)}_{y}\sigma^{(2)}_{x}].
\nonumber\\
\label{SH}
\end{eqnarray}
Note that we have $\Delta=0$ and $\theta=\pi/2$.
\item {\it Antisymmetric}: In this case we have  antiparallel magnetic moments, i.e., $\omega_+=-\omega_-=B$ and therefore 
\begin{eqnarray}
H_{\text{as}} & = & \frac{B}{2} [\sigma_z^{(1)} - \sigma_z^{(2)}] - 
J_{zz}\sigma^{(1)}_{z}\sigma^{(2)}_{z}
\nonumber\\
 & & + \frac{J}{2}[\sigma^{(1)}_{x}\sigma^{(2)}_{x}+\sigma^{(1)}_{y}\sigma^{(2)}_{y}] - 
\frac{D}{2}[\sigma^{(1)}_{x}\sigma^{(2)}_{y}-\sigma^{(1)}_{y}\sigma^{(2)}_{x}] 
\nonumber\\
 & & + \frac{r}{2}[\sigma^{(1)}_{x}\sigma^{(2)}_{x}-\sigma^{(1)}_{y}\sigma^{(2)}_{y}] + 
\frac{K}{2}[\sigma^{(1)}_{x}\sigma^{(2)}_{y}+\sigma^{(1)}_{y}\sigma^{(2)}_{x}], 
\nonumber\\. 
\label{ASH}
\end{eqnarray}
where in contrast to symmetric dimers we have $\omega=0$ and thus $\nu=\pi/2$.
\end{itemize}

\begin{figure}[h]
\begin{center}
\includegraphics[width=80mm]{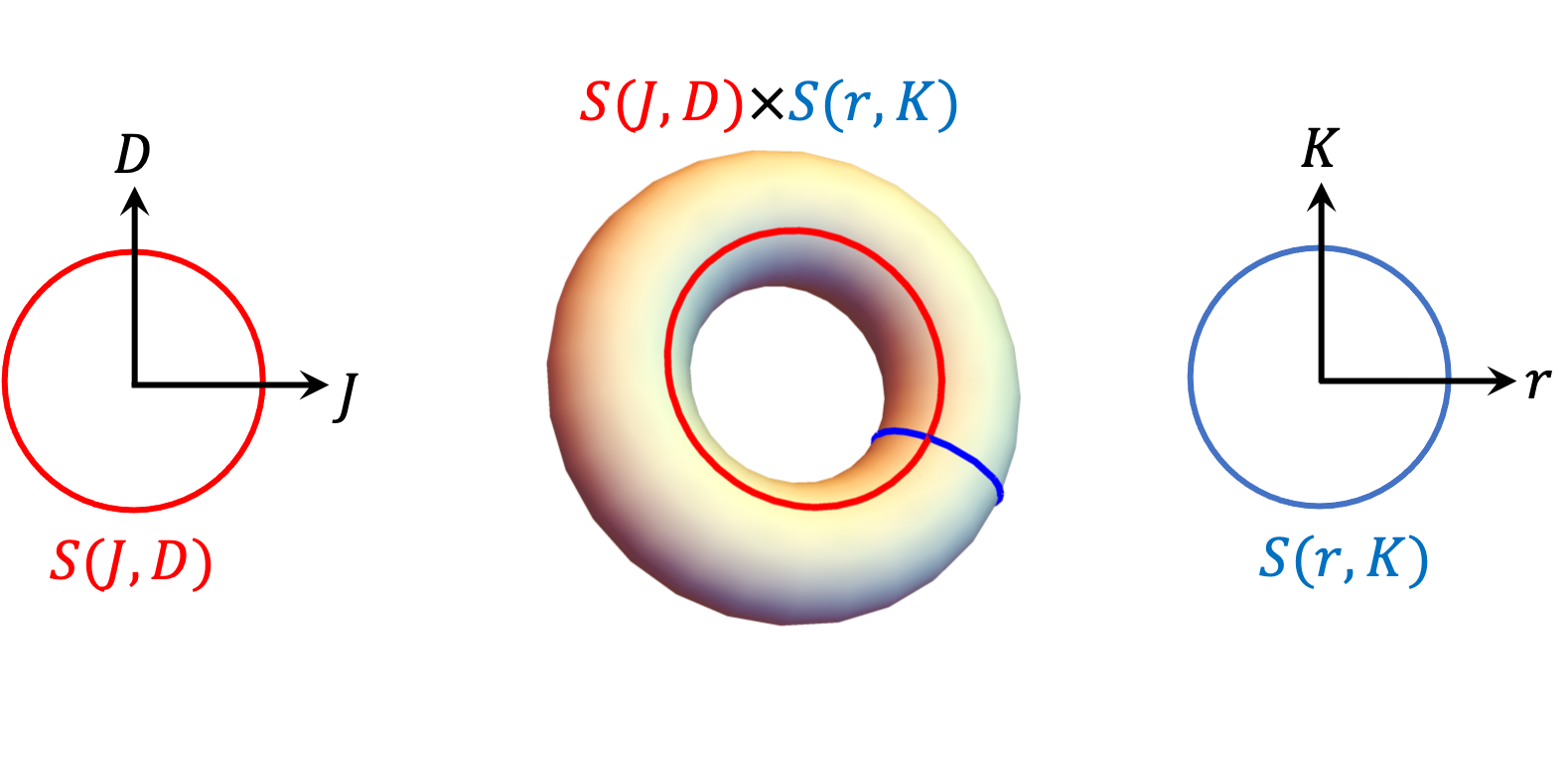}
\end{center}
\caption{(Color online) Schematic picture of a S-class or AS-class as a two dimensional 
torus in the $(J, D, r, K)$ parameter space.}
\label{fig:T1}
\end{figure}

We note that (I) and (II) in Eq.~\eqref{EDE} classify the above two categories of symmetric 
and antisymmetric dimers, each within its own category, into equivalence classes 
of dimers in a way that all the dimers in the same class have the same entanglement 
phase diagram in the temperature-magnetic field parameter space. Explicitly, if we solve 
Eq. \eqref{EDE} for the critical line in the $(B, T)$ plane, across which the thermal state 
changes its entangling feature from entangled to separable, then we find out that two 
dimers specified with interaction parameters $(J, D, r, K)$ and $(J', D', r', K)$ give rise 
to the same critical line in the $(B, T)$ plane or they are in the same class if and only if 
\begin{eqnarray}
\left\{ \begin{array}{ll} 
J^2+D^2=J'^2+D'^2 \\ 
r^2+K^2=r'^2+K'^2.
\end{array} \right. 
\label{ECWEC} 
\end{eqnarray}      
In other words, the above symmetric and antisymmetric categories consist of 
equivalence classes of dimers, where each class can be represented topologically as a 
two dimensional torus in the $(J, D, r, K)$ parameter space as shown in Fig. \ref{fig:T1}. 
Here, we may refer to an equivalence class of symmetric dimers as an S-class and an 
equivalence class of antisymmetric dimers as an AS-class. 

\begin{figure}[h]
\begin{center}
\includegraphics[width=80mm]{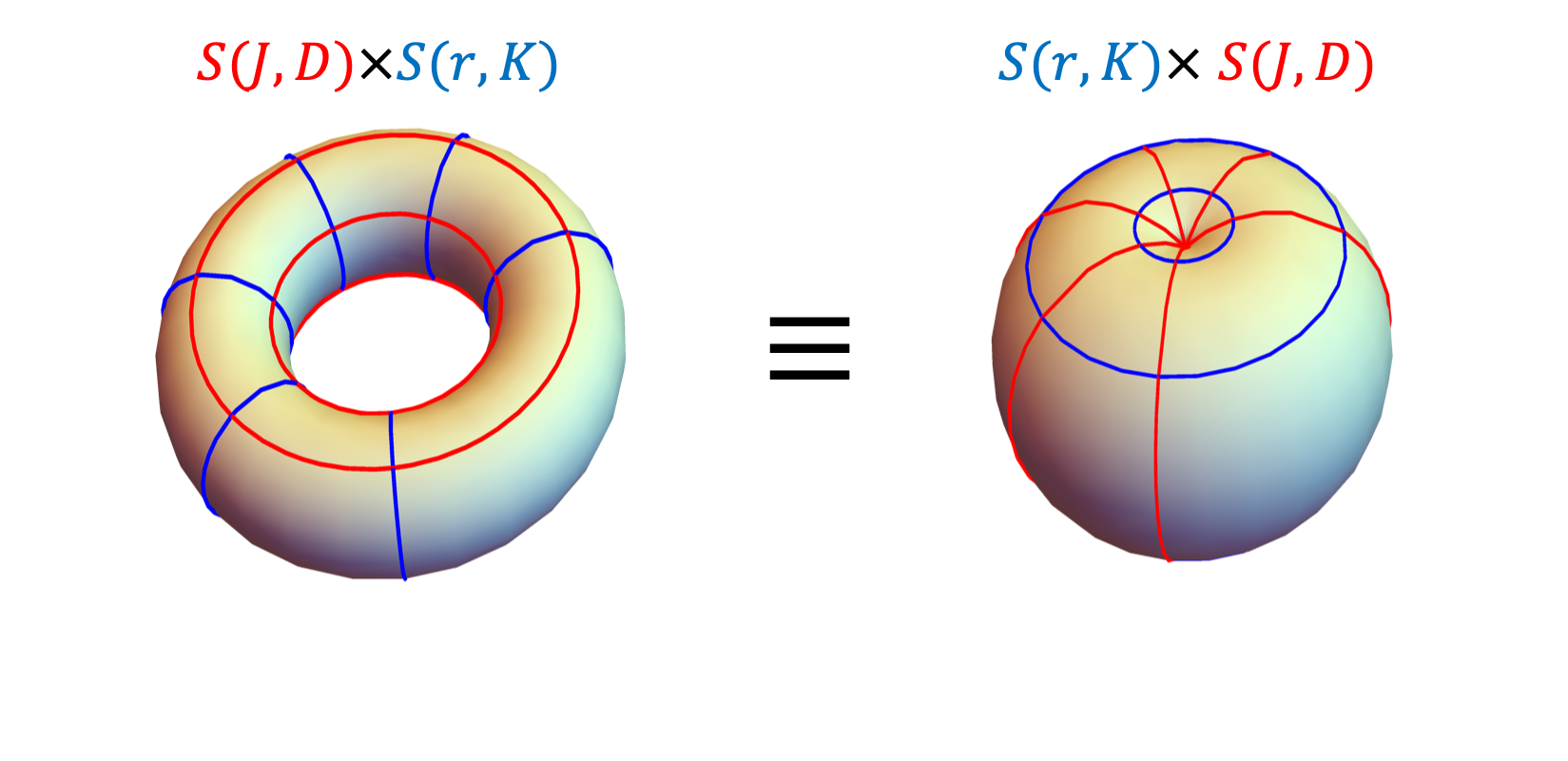}
\end{center}
\caption{(Color online) Toric visualisation of dual symmetric and antisymmetric entanglement 
classes. Each class has the same toric characteristic as its dual but with radial or meridian 
circles in swapped order.}
\label{fig:T2}
\end{figure}

Having each category of dimers classified into equivalent toric classes of entangled spins, 
we further notice that each class in one category has its one-to-one correspondence dual 
class in the opposite category. To see this we realize that the two equations in the duality 
equation Eq. \eqref{EDE} are indeed two sides of the same coin and actually are related 
by flipping the sign of the $J_{zz}$ coupling parameter and exchanging the $p$ and $q$ 
functions. Therefore, if a given toric S-class characterized by the coupling 
parameters $(J, D, r, K)$ is obtained by one of the equations in Eq. \eqref{EDE}, then its 
corresponding dual AS-class characterized by the coupling parameters 
$(\tilde{J}, \tilde{D}, \tilde{r}, \tilde{K})$ can be obtained from the other equation in 
Eq.~\eqref{EDE} and vice versa. The two dual toric S-class and AS-class are equivalent 
in a sense that they give rise to the same entanglement phase diagram in the $(B, T)$ 
plane if their characterisation parameters satisfy the following relations 
\begin{eqnarray}
\left\{ \begin{array}{ll} 
J^2+D^2=\tilde{r}^2+\tilde{K}^2 \\ 
r^2+K^2=\tilde{J}^2+\tilde{D}^2.
\end{array} \right. 
\end{eqnarray}  
These relations compared to the ones given in Eq. \eqref{ECWEC} indicate that the dual 
symmetric and antisymmetric entanglement classes have the same toric characteristic 
but with their radial or meridian circles in swapped order as depicted in Fig \ref{fig:T2}. 

Note that although the above analysis focuses on concurrence, we have 
found exactly the same classification and entanglement duality using non-locality \cite{horodecki1995,horst13} and negativity \cite{Vidal2002}. 
Other than the entanglement duality in spin-spin interactions demonstrated above, our 
analysis reveals the geometric and topological nature of quantum entanglement in a 
sense that entanglement phase diagram in dimers provide a clear geometric foliation 
of coupling parameter manifold into two dimensional compact torus leaves.

To further clarify our point and show that the entanglement analysis above provide a 
non-trivial classification and duality, we consider two examples in the following section. 

\section{Examples}

Consider two identical spin-$\frac{1}{2}$ particles with Heisenberg spin-spin exchange 
interaction of strength $J$ in an external magnetic field, as given by the Hamiltonian    
\begin{eqnarray}
H^{\rm{Heisenberg}} & = &B\left( s_z^{(1)} + s_z^{(2)} \right) + J {\bf s}^{(1)} \cdot {\bf s}^{(2)} . 
\label{eq:Heisenberg}
\end{eqnarray} 
Here, ${\bf s}=\frac{1}{2}\boldsymbol{\sigma}$ and $B$ is the homogeneous field strength. 
We assume antiferromagnetic coupling $J>0$ as entanglement cannot occur for ferromagnetic 
coupling \cite{arnesen01}. 

According to the above classification the Heisenberg model describes a class of symmetric 
dimers, which includes all the spin-spin interaction models following the general form of 
the Hamiltonian
\begin{eqnarray}
H^{\rm{Heisenberg}}_{\text{s}} & = & B\left( s_z^{(1)} + s_z^{(2)} \right) + Js^{(1)}_{z}s^{(2)}_{z} 
\nonumber \\
 & & + J'[s^{(1)}_{x}s^{(2)}_{x}+s^{(1)}_{y}s^{(2)}_{y}] 
\nonumber \\
 & &- D'[s^{(1)}_{x}s^{(2)}_{y}-s^{(1)}_{y}s^{(2)}_{x}] 
 \nonumber \\
\label{eq:GHeisenberg}
\end{eqnarray} 
such that $J'^2+D'^2=J^2$. Note that $H^{\rm{Heisenberg}}_{\text{s}}=H_{\rm{Heisenberg}}$ for $D'=0$ and $J'=J$. 
Below we show all dimers in this class have the same entanglement phase diagram in 
$(B, T)$ plane.  

\begin{figure}[h]
\begin{center}
\includegraphics[width=85mm]{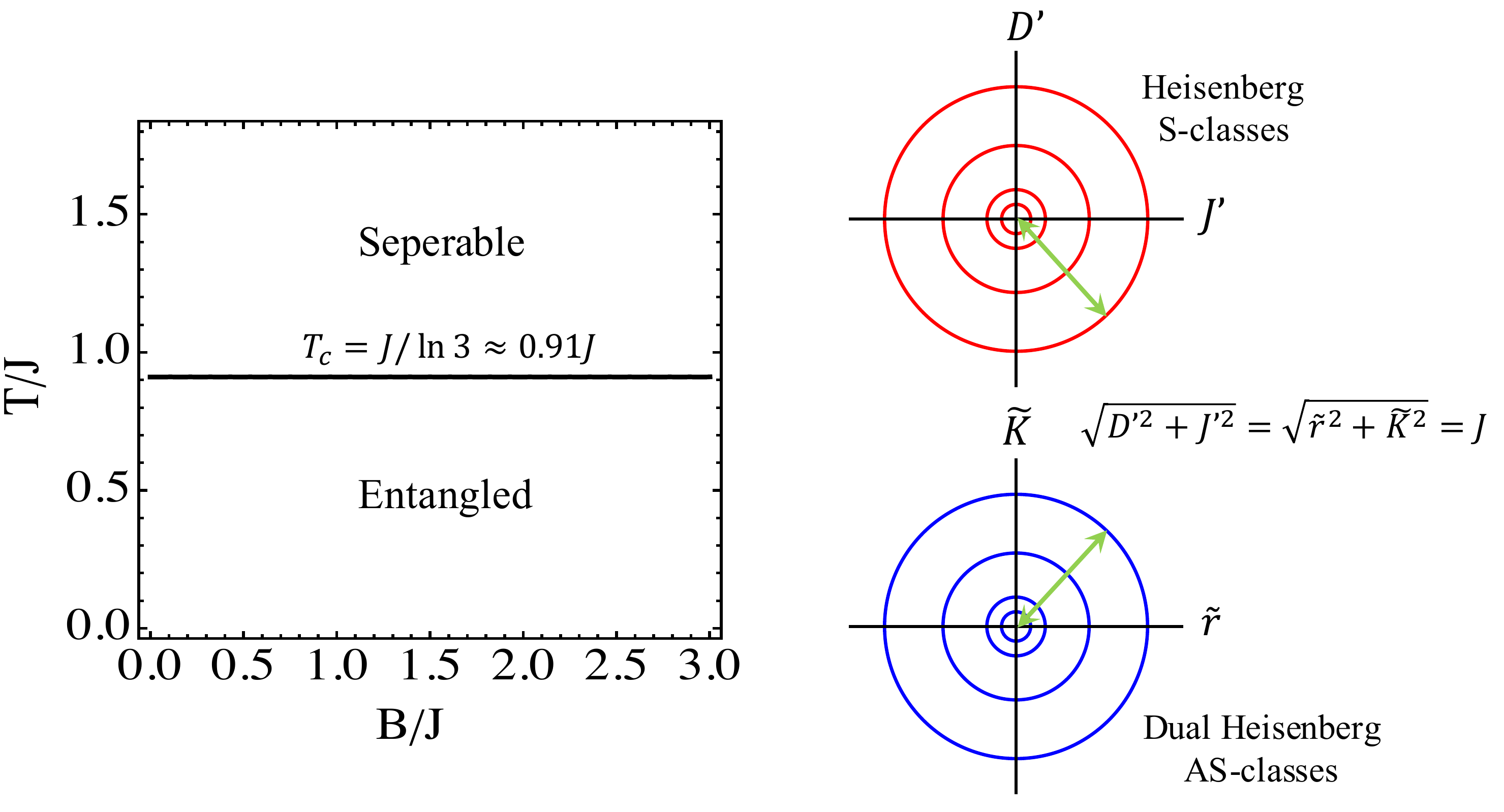}
\end{center}
\caption{(Color online) Left panel shows the entanglement phase diagram for symmetric 
and dual antisymmetric Heisenberg dimers. For each of the Heisenberg type of dimers, 
the entanglement undergoes a sudden change at the relative critical temperature 
$T_{c} =T'_{c}=\tilde{T_{c}}= J/\ln 3 \approx 0.91 J$ independent of the applied magnetic 
field strength $B$, which is an indication of quantum phase transition. Right panels illustrate 
toric visualisation of symmetric (S) and dual antisymmetric (AS) Heisenberg entanglement 
classes. While each circle of radius $J$ in the $(J', D')$ plane represent a Heisenberg 
S-class corresponding to a given isotropic Heisenberg exchange coupling $J$, the equivalent 
dual AS-class is represented by a circle in the $(\tilde{r}, \tilde{K})$ plane with the same 
radius of $J$.}
\label{fig:HCandPD}
\end{figure}

For the Hamiltonian in Eq. \eqref{eq:GHeisenberg}, the thermal state is characterized by 
the following nonzero elements:  
\begin{eqnarray}
\rho'_{11} & = & \frac{1}{\mathcal{Z}} e^{-B/T}, \ \  \rho'_{44} = \frac{1}{\mathcal{Z}'} e^{B/T} ,  
\nonumber \\ 
\rho'_{22} & = & \rho'_{33} = \frac{e^{J/2T}}{\mathcal{Z}'} \cosh\left(\Gamma'/2T\right) , 
\nonumber \\ 
\rho'_{23} & = & (\rho'_{32})^{\ast} = -\frac{e^{J/2T}}{\mathcal{Z}'} e^{-i\phi}\sinh\left(\Gamma'/2T\right) 
\end{eqnarray} 
with partition function
\begin{eqnarray}
\mathcal{Z}' &=&  2e^{J/2T} \cosh\left(\Gamma'/2T\right)  + 2\cosh \left(B /T \right),
\end{eqnarray}
where $\Gamma'=\sqrt{J'^2+D'^2}=J$. The concurrence functions are then given by
\begin{eqnarray}
C_{1}&=&-\frac{e^{J/2T}}{\mathcal{Z}'} \cosh\left(\Gamma'/T\right) = 
-\frac{ \left( 1+e^{J/T}\right)}{2\mathcal{Z}'}< 0,
\nonumber \\
C_{2} & = & \frac{1}{\mathcal{Z}}\left[e^{J/2T}\sinh\left(\Gamma'/T\right)-1\right] = 
\frac{e^{J/T}-3}{2\mathcal{Z}'}
\label{Duality-Eq-S}
\end{eqnarray}
which result in
\begin{eqnarray}
C(\varrho'_T) =\max \left\{ \frac{e^{J/T}-3}{\mathcal{Z}'} , 0 \right\} . 
\label{C-HH}
\end{eqnarray}
Thus, $\varrho'_T$ is entangled if $e^{J/T}-3 >0$. This defines a critical 
temperature
\begin{eqnarray}
T'_{c} = \sqrt{J'^2+D'^2}/ \ln 3 = J/\ln 3 \approx 0.91 J,
\label{eq:critical_hm_entanglement}
\end{eqnarray}
which is independent of the external magnetic field $B$, for all dimers in the Heisenberg 
class denoted here by $[H^{\rm{Heisenberg}}]$. As shown in Fig. \ref{fig:HCandPD}, above 
the critical temperature the thermal state ceases to be entangled. The same critical 
temperature as in Eq. \eqref{eq:critical_hm_entanglement} has been obtained in 
Ref.~\cite{arnesen01} for antiferromagnetic Heisenberg exchange without the 
Dzyaloshinskii–Moriya interaction, i.e., $D'=0$. Note that the toric characteristic 
equation 
\begin{eqnarray}
J'^2+D'^2=J^2
\end{eqnarray}
identifies each symmetric Heisenberg entanglement class as a one dimensional torus 
(one-sphere) in the $(J', D)$ plane with the radius given by the strength of the Heisenberg 
coupling constant $J$. In other words, symmetric Heisenberg entanglement classes 
foliates the $(J', D')$ parameter space into circles (see Fig. \ref{fig:HCandPD}).

On the opposite category of antisymmetric dimers, the dual Heisenberg model is described 
by the antisymmetric Hamiltonian 
\begin{eqnarray} 
H^{\rm{Heisenberg}}_{\text{as}} & = &B\left( s_z^{(1)} - s_z^{(2)} \right) - Js^{(1)}_{z}s^{(2)}_{z} \nonumber \\ 
 & & + \tilde{r}[s^{(1)}_{x} s^{(2)}_{x}-s^{(1)}_{y}s^{(2)}_{y}] 
 \nonumber \\
 & &+ \tilde{K}[s^{(1)}_{x}s^{(2)}_{y} + s^{(1)}_{y}s^{(2)}_{x}],
\nonumber\\ 
\label{eq:DHeisenberg} 
\end{eqnarray} 
where 
\begin{eqnarray}
\tilde{r}^2+\tilde{K}^2=J^2
\label{DCE}
\end{eqnarray}
with $J$ being the Heisenberg coupling constant in Eq. \eqref{eq:Heisenberg}. In this case, 
the dual thermal state is given by the nonzero elements 
\begin{eqnarray}
\tilde{\rho}_{11} & = &\tilde{\rho}_{44} = \frac{e^{J/2T}}{\tilde{\mathcal{Z}}} 
\cosh \left( \tilde{\Gamma} /2T \right)  ,  
\nonumber \\ 
\tilde{\rho}_{22} & = &  \frac{1}{\tilde{\mathcal{Z}}} e^{-B/T},\ \ \tilde{\rho}_{33} = 
\frac{1}{\tilde{\mathcal{Z}}} e^{B/T}, 
\nonumber \\ 
\tilde{\rho}_{14} & = & \tilde{\rho}_{41}^{\ast} = -\frac{e^{J/2T}}{\tilde{\mathcal{Z}}} 
e^{-i\varphi}\sinh\left(\tilde{\Gamma}/2T\right) 
\end{eqnarray} 
with partition function
\begin{eqnarray}
\tilde{\mathcal{Z}} &=&  2e^{J/2T} \cosh\left(\tilde{\Gamma}/T\right)  + 2\cosh \left(B /T \right),
\end{eqnarray}
where $\tilde{\Gamma}=\sqrt{\tilde{r}^2+\tilde{K}^2}=J$. The concurrence functions are given by
\begin{eqnarray}
C_{1} & = & \frac{1}{\tilde{\mathcal{Z}}}\left[e^{J/2T}\sinh\left(\tilde{\Gamma}/T\right)-1\right] = 
\frac{e^{J/T}-3}{2\tilde{\mathcal{Z}}},
\nonumber \\
C_{2} & = & -\frac{e^{J/2T}}{\tilde{\mathcal{Z}}} \cosh\left(\tilde{\Gamma}/T\right) = 
-\frac{ \left( 1+e^{J/T}\right)}{2\tilde{\mathcal{Z}}}< 0,
\end{eqnarray}
which compared to its symmetric counterparts in Eq. \eqref{Duality-Eq-S} the $C_{1}$ 
and $C_{1}$ functions are swapped.
This results in
\begin{eqnarray}
C(\tilde{\varrho}_T) =\max \left\{ \frac{e^{J/T}-3}{\tilde{\mathcal{Z}}} , 0 \right\} . 
\label{C-HH}
\end{eqnarray}

Similar to the symmetric Heisenberg dimers, the dual thermal state $\tilde{\varrho}_T$ is 
also entangled if $e^{J/T}-3 >0$. This indicates the entanglement phase diagram with the  
critical temperature
\begin{eqnarray}
\tilde{T}_{c} = \sqrt{\tilde{r}^2+\tilde{K}^2}/ \ln 3 = J/\ln 3 \approx 0.91 J,
\end{eqnarray}
for antisymmetric Heisenberg dimers to be the same as one obtained for symmetric 
Heisenberg dimers above. However, in the dual antisymmetric case the toric characteristic 
equation given by Eq. \eqref{DCE}, as shown in Fig. \ref{fig:HCandPD} foliates instead 
the $(\tilde{r}, \tilde{K})$ parameter space into circles of radii specified by the 
Heisenberg exchange parameter $J$.

\begin{figure}[h]
\begin{center}
\includegraphics[width=85mm]{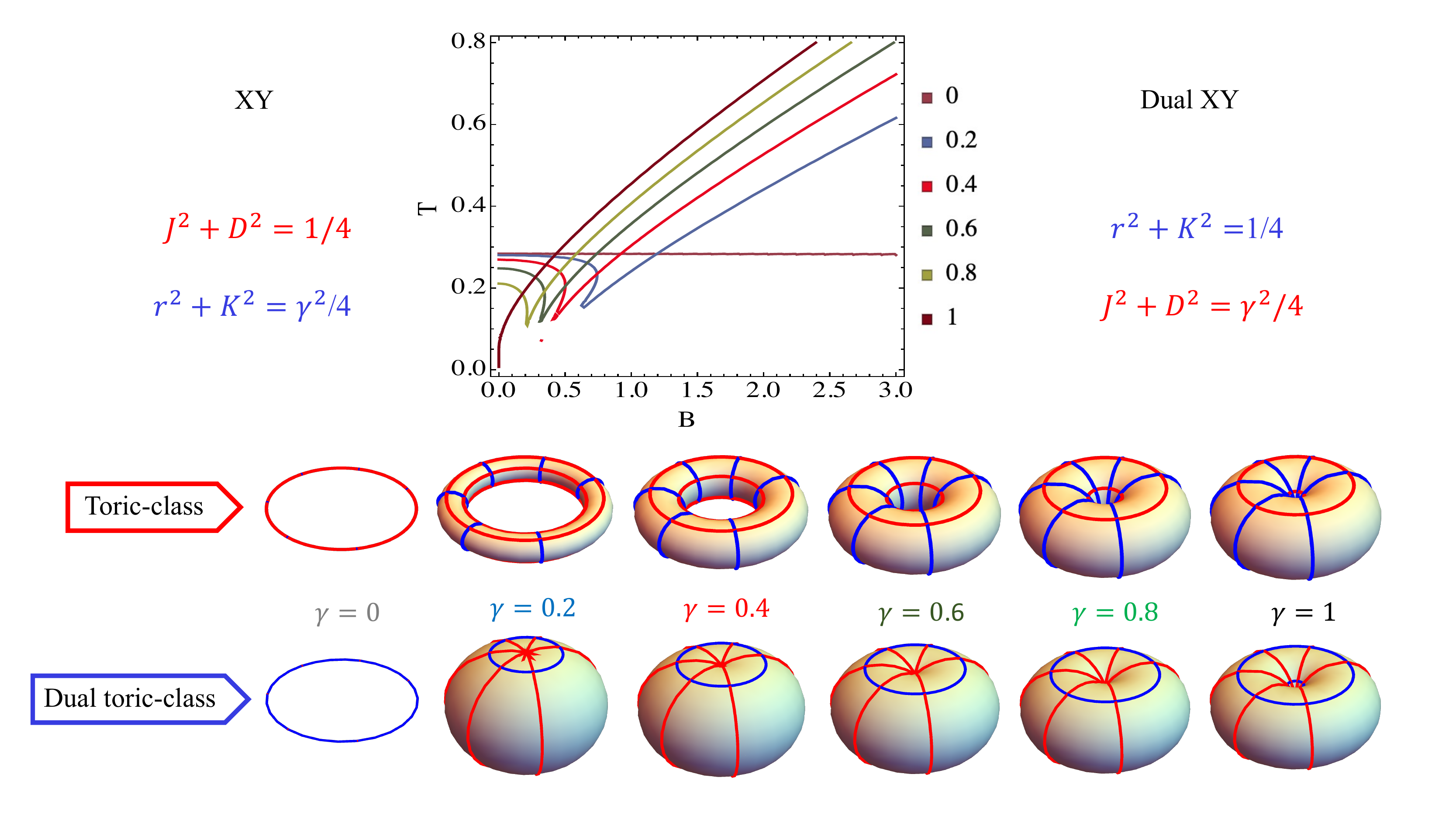}
\end{center}
\caption{(Color online) Upper panel shows the entanglement transition curve for symmetric 
and dual antisymmetric XY dimers for different values of anisotropy parameter $\gamma$. While the isotropic limit $\gamma= 0$ represents the 
narrowest area of entanglement with magnetic field independent critical temperature, the critical temperature becomes magnetic field dependent for $\gamma\neq0$ and the entanglement area increases with $\gamma$
so that the Ising limit $\gamma = 1$ represents the broadest entanglement area in the temperature-magnetic field plane. 
Each value of $\gamma$ distinguishes exclusive dual toric entanglement classes depicted in the lower panel.}
\label{fig:XY}
\end{figure}

In addition, Fig. \ref{fig:XY} illustrates the classification of the XY interaction, given by the Hamiltonian  
\begin{eqnarray}
H^{\rm XY}_{\text{s/as}} &=& (1+\gamma)s_x^{(1)} s_x^{(2)} + (1-\gamma) s_y^{(1)} s_y^{(2)}
\nonumber\\
 & & + B (s_z^{(1)}\pm s_z^{(2)}),
\label{eq:xy}
\end{eqnarray}  
into dual toric entanglement classes. Here  $\gamma \in [0,1]$ is the dimensionless 
anisotropy parameter controlling the cylindrical asymmetry of the spin-spin interaction and 
$+$ ($-$) is for symmetric (antisymmetric) XY model. 
The XY model defines interactions from the isotropic limit $\gamma= 0$ with additional 
symmetry $[H^{\rm XY}_{\text{s/as}}, s^z] = 0$ to the opposite limit $\gamma = 1$, which corresponds to the 
Ising model with totally ordered Neel ground state. By comparing the Hamiltonian in 
Eq.~\eqref{eq:xy} with the dual symmetric and antisymmetric Hamiltonians in Eqs.~\eqref{SH} 
and \eqref{ASH} for $J_{zz}=0$ and then applying the above classification procedure to 
the XY model, we fine dual characteristic equations as
\begin{eqnarray}
\left\{ \begin{array}{ll} 
J^{2} +D^{2}=1/4 \\ 
r^{2}+K^{2}=\gamma^{2}/4 
\end{array} \right. \equiv
\left\{ \begin{array}{ll} 
J^{2} +D^{2}=\gamma^{2}/4 \\ 
r^{2}+K^{2}= 1/4
\end{array} \right.
\end{eqnarray} 
which specify dual toric XY entanglement classes as depicted in the lower panel of 
Fig.~\ref{fig:XY}. In fact, it is the anisotropy parameter $\gamma$ that controls the 
entanglement classes and the spin-spin duality. Note that 
the critical temperature depends on applied magnetic field for all $\gamma\neq 0$.
 
\section{Conclusions}
In conclusion, we have demonstrated an entanglement duality of a wide class of 
physically important spin-spin interaction dimer models by analyzing entanglement 
transition of thermal states in the temperature-magnetic field plane. This duality allows 
us to foliate coupling parameter space into dual symmetric and antisymmetric toric 
entanglement classes. This classification is an indication of topological nature of 
the quantum correlations and hope it can contribute to our deeper 
understanding of the various aspects of the mysterious concept of the quantum 
correlations.

\section{Acknowledgment}
The authors acknowledge financial support from the Knut and Alice Wallenberg Foundation 
through Grant No. 2018.0060. E.S. acknowledges financial support from the Swedish Research 
Council (VR) through Grant No. 2017-03832.

\end{document}